\documentstyle[aps,prl]{revtex}
\begin{document}
\input epsf.sty
\twocolumn[\hsize\textwidth\columnwidth\hsize\csname %
@twocolumnfalse\endcsname
\draft
\widetext
%%%%%%%% prl (above) %%%%%%%%%%%%%%%%%%

\title{Magnetic-field-induced collapse of charge-ordered nanoclusters 
and the Colossal 
Magnetoresistance effect in Nd$_{0.7}$Sr$_{0.3}$MnO$_3$}

\author{T. Y. Koo$^1$, V. Kiryukhin$^1$, P. A. Sharma$^1$, 
 J. P. Hill$^2$, and S-W. Cheong$^{1,3}$}
\address{(1) Department of Physics and Astronomy, Rutgers University,
Piscataway, New Jersey 08854}
\address{(2) Department of Physics, Brookhaven National Laboratory, Upton,
New York 11973}
\address{(3) Bell Laboratories, Lucent Technologies, Murray Hill,
New Jersey 07974}

\date{\today}
\maketitle

\begin{abstract}
We report synchrotron x-ray scattering studies of
charge/orbitally ordered (COO) nanoclusters
in Nd$_{0.7}$Sr$_{0.3}$MnO$_3$. 
We find that the COO nanoclusters are strongly suppressed in an applied
magnetic field, and that their decreasing concentration follows the
field-induced decrease of the sample electrical resistivity. 
The COO nanoclusters, however, do not completely disappear in the conducting
state, suggesting that this state is inhomogeneous and contains an admixture
of an insulating phase. 
Similar results were also obtained for the zero-field insulator-metal
transition that occurs as temperature is reduced. 
These observations suggest that these correlated
lattice distortions play a key role in the Colossal Magnetoresistance effect 
in this prototypical manganite.

\end{abstract}


%%%%%% prl format (below) %%%%%%%%%%%%%%
\phantom{.}
]
\narrowtext
%%%%%%%%% prl (above) %%%%%%%%%%%%%%%%%%

Manganite perovskites of the chemical formula A$_{1-x}$B$_x$MnO$_3$ (where
A is a rare earth, and B is an alkali earth atom) have recently attracted 
considerable attention because they exhibit a number of interesting electronic
properties \cite{Review}. Perhaps the most dramatic of these is the
magnetic-field-induced insulator-metal 
transition, often referred to as the Colossal
Magnetoresistance (CMR) effect. In its most widely studied form, the CMR
effect is a transition from a paramagnetic insulating (PI) to a ferromagnetic
metallic (FM) phase. The very large difference between the electrical 
resistivities of the PI and FM phases lies at the core of the CMR effect, and
considerable efforts have been spent in order to explain this difference
\cite{Review}. The 
metallic nature of the FM phase was first explained in the 1950's in the
framework of the double-exchange mechanism \cite{Zener}. 
In this model, itinerant 
$e_g$ electrons have their spins aligned with the localized $t_{2g}$ core spins
of the Mn atoms by virtue of a strong Hund coupling, and the $e_g$ electrons
can therefore hop easily between the ferromagnetically aligned Mn atoms.
Conversely, the motion of the $e_g$ electrons is suppressed in the 
disordered paramagnetic environment, and the PI phase should, therefore, exhibit
larger resistivity.

However, in the 1990's it was pointed out that the change in resistivity was far
too large to be explained by the double exchange model \cite{Millis}.
Theoretical \cite{Millis,Brat}
and experimental \cite{Polarons,PDF,xrayPolarons} work has suggested that 
the high resistivity of the PI phase results in
part from the presence of the lattice polarons that form when an $e_g$ electron
localizes on a Mn$^{3+}$ site, inducing a Jahn-Teller
distortion of the MnO$_6$ octahedron.

At low temperatures,
the CMR manganites exhibit a variety of charge and orbitally ordered phases.
Further,
charge and orbital correlations are in many cases found even in the samples 
which do not show
the corresponding long-range order, often giving rise to microscopically
inhomogeneous states \cite{D,Other,Corr,Corr1}. In fact,
short-range structural correlations were recently observed in the PI phase in
La$_{1-x}$Ca$_x$MnO$_3$, La$_{2-2x}$Sr$_{1+2x}$Mn$_2$O$_7$, and
Pr$_{1-x}$Ca$_x$MnO$_3$ manganites \cite{Corr,Corr1,xrayPolarons}.
These correlations are interpreted as arising from nanoscale
regions possessing charge and orbital order and 
associated Jahn-Teller lattice distortions.
The latter distortions form a periodic lattice modulation in the
ordered regions, making it possible to detect these regions using x-ray
and neutron scattering techniques.
Because the characteristic correlation length
of these regions is typically 10-20 $\rm\AA$, 
we refer to them as charge/orbitally
ordered (COO) nanoclusters.
The electrical resistivity has been shown to
increase with the concentration of these correlated
regions \cite{xrayPolarons,Corr}, indicating that they are insulating. 
Thus, in addition to the single polarons discussed above, correlated regions can
contribute to the high
resistivity of the PI phase.

The exact role of the COO nanoclusters
in the magnetic-field induced insulator-metal
transition (the CMR effect) is, however,
currently not clear. In this work, we study
the effects of a magnetic field on these structural correlations 
in a single crystal of Nd$_{0.7}$Sr$_{0.3}$MnO$_3$ using x-ray diffraction. 
We find that as the sample
undergoes a magnetic-field-induced insulator-metal transition, 
the structural correlations due to the COO nanoclusters
are strongly suppressed. The COO nanoclusters, however, do not disappear 
completely in the conducting
state, suggesting that this state consists of a mixture of metallic and
insulating regions. Similar results were also obtained
for the insulator-metal transition that occurs as the temperature is reduced in
zero magnetic field. 
These observations suggest that the  correlated
lattice distortions play a key role in insulator-metal transitions 
in this prototypical manganite.

Single crystals of Nd$_{0.7}$Sr$_{0.3}$MnO$_3$ were grown using the standard
floating zone technique. The x-ray diffraction measurements were carried out
at beamlines X22B and X22C at the National Synchrotron Light Source. 
The x-ray beam was focused by a mirror, monochromatized by a Ge (111)
monochromator, scattered from the sample,
and analyzed with a pyrolytic graphite crystal. The sample was mounted either
in a closed-cycle refrigerator (T=10-450 K), or in a 13 Tesla superconducting
magnet. In the former case, the x-ray energy was 10 keV and a
vertical scattering geometry was employed, while in the latter
7.8 keV x-rays and a horizontal scattering geometry were utilized. 
In this paper,
Bragg peaks are indexed in the orthorhombic {\it Pbnm}
notation in which the longest
lattice constant is $c$. The scattering vectors ($h, k, l$) are given in
reciprocal lattice units.

\begin{figure}
\centerline{\epsfxsize=2.9in\epsfbox{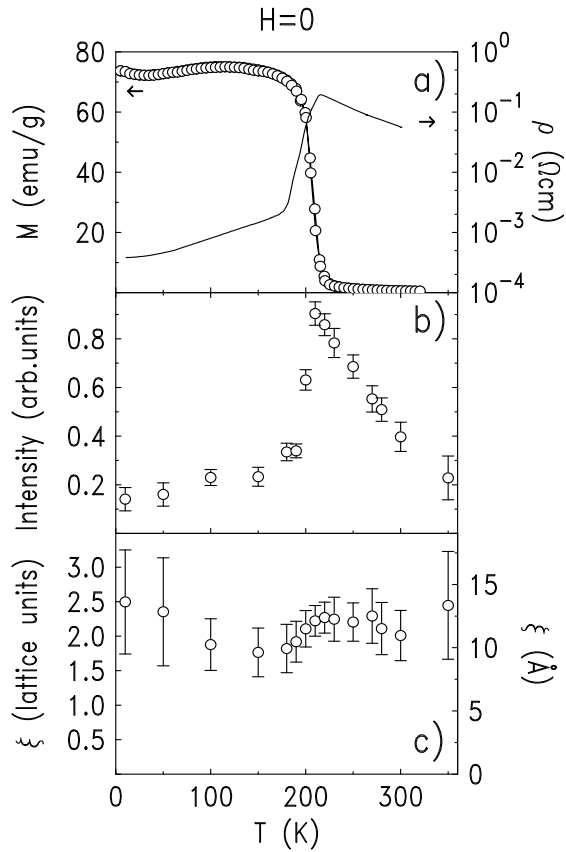}}
\vskip 5mm
\caption{(a) Temperature dependence of the zero-field electrical resistivity 
(solid line) and
the magnetization (open circles) in a
magnetic field of 2000 Oe of Nd$_{0.7}$Sr$_{0.3}$MnO$_3$. (b) The
intensity of the (4, 4.5, 0) peak due to the correlated structural
distortions. The
single-polaron background is subtracted as discussed in the text.
(c) The correlation length of the ordered regions.}
\label{fig1}
\end{figure}

\begin{figure}
\centerline{\epsfxsize=2.9in\epsfbox{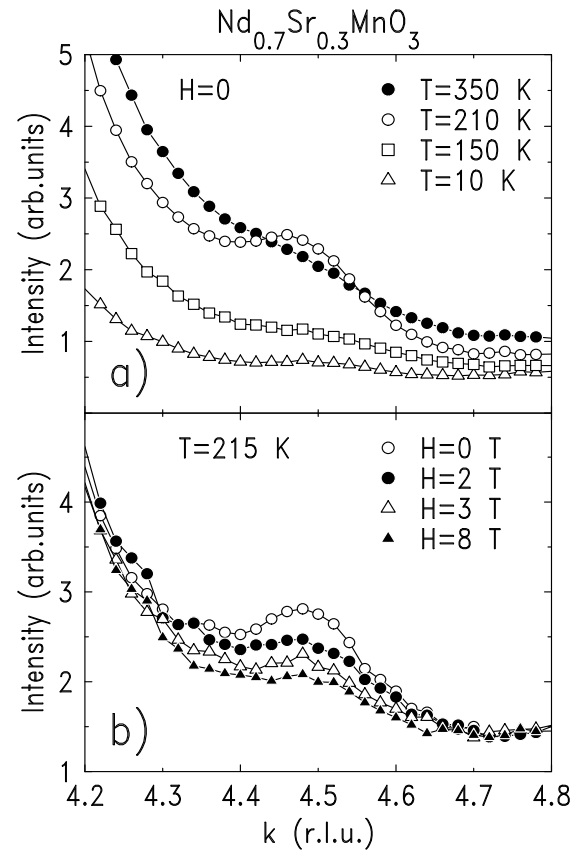}}
\vskip 5mm
\caption{X-ray scans along the (4, k, 0) direction taken in (a) zero magnetic
field and various temperatures, and (b) T=215 K and various fields.}
\label{fig2}
\end{figure}

Before presenting the field dependent data, we first discuss the zero-field
behavior of Nd$_{0.7}$Sr$_{0.3}$MnO$_3$. 
This compound is a paramagnetic insulator at high temperatures.
With decreasing temperature, it undergoes a transition to a ferromagnetic 
metallic state at T$_c$$\approx$210 K, see Fig. \ref{fig1}(a).
The properties of this compound are, therefore, similar to those of 
La$_{0.7}$Ca$_{0.3}$MnO$_3$. In the latter compound, the PI state was found
to exhibit short-range structural correlations with the reduced wave vector
of ($h, k, l$)=(0, 0.5, 0) 
and a correlation length of several lattice constants \cite{Corr,Corr1}.
We find that Nd$_{0.7}$Sr$_{0.3}$MnO$_3$ exhibits the same kind of structural
correlations. Fig. \ref{fig2}(a) shows scans along the (4, $k$, 0) direction
in the reciprocal space. A clear peak is seen at $k$=4.5 at T=210 K, just
above T$_c$. The peak is observed on a sloping background which is
attributed to the scattering due to single polarons \cite{xrayPolarons}, 
also known as Huang scattering, and to thermal-diffuse scattering.
Similar data were taken at a number of temperatures. The results were fitted
to a sum of a peak with
the Lorentzian-square line shape, and a monotonically sloping
background, the latter
described by a power-law function. The intensity of the
peak is proportional to both the concentration of the correlated regions and
to the square of the magnitude of the lattice distortion in these regions.
However, neutron measurements of the pair-distribution function (PDF) in related
samples indicate that the magnitude of the lattice distortion does not
change with temperature \cite{PDF}. Therefore,
we interpret the temperature dependence of this peak 
as being simply proportional to the temperature dependence of
the concentration of the correlated regions in the sample.
The width of the peak is inversely proportional to the correlation length of
these regions. 

Fig. \ref{fig1}(b) shows the intensity of the (4, 4.5, 0) peak as a function
of temperature. As observed in La$_{0.7}$Ca$_{0.3}$MnO$_3$
\cite{Corr}, this intensity
always decreases with decreasing resistivity [Fig. \ref{fig1}(a)]. This behavior
is consistent with a picture in which the $e_g$ electrons in the correlated
regions are localized: as the nanoclusters
disappear, the resistivity decreases.
Interestingly, the data of Fig. \ref{fig1}(b) show that
the lattice correlations do not disappear even
at the lowest temperatures, deep in the FM phase. 
The FM phase, therefore, must be inhomogeneous,
always containing insulating COO nanoclusters. 
Hence, it appears that the insulator-metal transition in this compound
is a complex process in which the insulating phase gradually disappears as the
volume fraction of the metallic phase grows. 
One possible description of this process can be given in terms of percolative
two-fluid model discussed in Ref. \cite{Jaime}.
Similar descriptions 
based on the results of small-angle neutron scattering and tunneling
spectroscopy measurements were also 
proposed in Refs. \cite{DeT,Mydosh}.
Our observations are also
consistent with the results of the neutron PDF measurements
in related samples \cite{PDF}. The latter 
measurements showed that local
Jahn-Teller lattice
distortions, albeit at a small concentration, 
are present in the FM phase. Our data demonstrate
that at least some of these distortions stem from the COO nanoclusters.
 
Fig. \ref{fig1}(c) shows the correlation length of the ordered nanoclusters, 
which is defined as the inverse half-width-at-half-maximum of
the (4, 4.5, 0) peak. This correlation length is between 2 and 3 lattice
constants and does not depend on temperature, to within errors.
The above definition would give the correct correlation length for exponentially
decaying correlations and a Lorentzian line shape. Note, however,
the actual size of the correlated regions depends on the
precise form of the correlation function -- which is currently unknown.
For example,
simple calculations of the structure factor for perfectly ordered clusters of a
finite size show that in such a case
the correlation length, as defined above, would underestimate
the cluster size by as much as a factor of two.

\begin{figure}
\centerline{\epsfxsize=2.9in\epsfbox{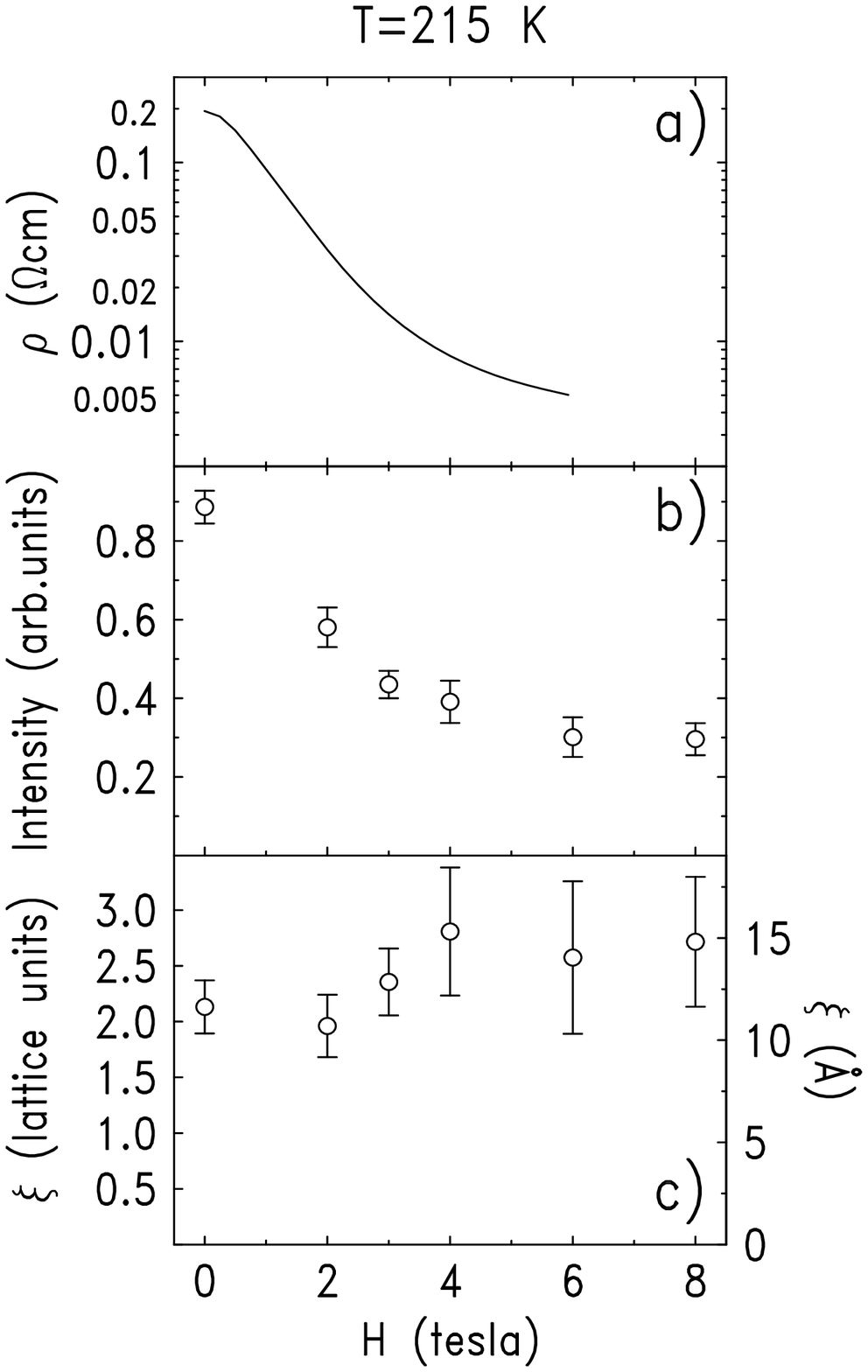}}
\vskip 5mm
\caption{Magnetic field dependence of (a) the electrical resistivity, (b)
the intensity of the (4, 4.5, 0) peak due to the correlated distortions, and
(c) the correlation length of the ordered regions.
All the data were taken at T=215 K.}
\label{fig3}
\end{figure}

We now turn to the main question of this paper--the role
of the COO nanoclusters in the CMR phenomenon.
To study the effects of a magnetic field on the nanoclusters, x-ray
experiments were carried out in a magnetic field at T=215 K, 
just above the Curie temperature. Figure \ref{fig3}(a) illustrates that
at T=215 K, the application of a magnetic field results in a  
substantial drop of the electrical resistivity.
X-ray scans at the (4, 4.5, 0) peak position taken at this 
temperature in various magnetic fields are shown in Fig. \ref{fig2}(b).
The intensity of the correlated peak arising from the COO nanoclusters
clearly diminishes 
with increasing field.
However, the correlations remain present in the
material even at the highest fields, deep in the metallic phase, as was
also the case at low temperatures in the
FM phase in zero field. Therefore, as in the latter case, 
the magnetic-field-induced conducting state is inhomogeneous, and the
field-induced insulator-metal transition may proceed via a complex process
involving coexistence of conducting and insulating phases.

The data of Fig.
\ref{fig2}(b) were analyzed in the same manner as the temperature-dependent
data. The fitted intensity of the correlated peak and the correlation length of
the ordered nanoclusters
are shown in Figs. \ref{fig3}(b), and (c), respectively. 
The correlated regions exhibit a correlation length
of 2-3 lattice constants, 
and their concentration, which is proportional to the amplitude of the
(4, 4.5, 0) peak, decreases with the increasing field, tracing the behavior
of the resistivity. 
Interestingly,
the size of the correlated regions does not change
as the magnetic field is varied, to within errors.
As discussed above, this size also does not depend on
temperature in zero magnetic field.
Further, this same cluster size was
previously obtained for La$_{0.7}$Ca$_{0.3}$MnO$_3$ samples \cite{Corr1}.
These observations suggest that there is an intrinsic mechanism defining
the size of the nanoclusters common to all these materials.

To explain the effect of the magnetic field on the correlated lattice
distortions, we first need to understand 
crystallographic and magnetic structure of
the COO nanoclusters.
Several possible model structures describing the observed lattice correlations
have been proposed. In one model \cite{KHKim,Corr,Corr1},
the correlated domains are described as
small regions possessing the CE-type charge and orbital ordering with its
checker-board-type charge order and the characteristic
orbital ordering \cite{Review}. This
picture is based on the observed lattice modulation
wave vector, which is the same for the long-range ordered CE-type structure
and for the short-range correlations in La$_{0.7}$Ca$_{0.3}$MnO$_3$ and
Nd$_{0.7}$Sr$_{0.3}$MnO$_3$. In addition, in Pr$_{0.7}$Ca$_{0.3}$MnO$_3$ these
correlations were observed to directly evolve into large regions
possessing the CE-type order. A uniform
magnetic field is known to destroy CE-type order \cite{Review}, and
hence such a model is consistent with our measurements.
Other descriptions of the correlated regions, such as correlated polarons,
and a bipolaron model, have also been proposed \cite{Corr,Corr1}.
In addition, in layered manganites, short-range
structural correlations were observed at a different wave vector. These
correlations were described as a charge density wave arising from
nesting properties of the 
Fermi surface \cite{Chu}.
We are currently unaware of any predictions for the effect of a
magnetic field in these
models.

The data of Figs. \ref{fig1}-\ref{fig3} show that 
the COO nanoclusters are suppressed in the ferromagnetic metallic phase 
independent of whether the insulator-metal transition is induced with a
magnetic field or by changing the temperature. Together with the observed
correlation between the concentration of the nanoclusters and the electrical
resistivity, these data
strongly suggest that the nanoclusters play a key role in
insulator-metal transitions in Nd$_{0.7}$Sr$_{0.3}$MnO$_3$.

Finally, we note that, in contrast to the intensity of the (4, 4.5, 0) peak, the
intensity of the sloping background in Fig. \ref{fig2}(b) does not show any
significant change as a function of applied magnetic field ($\Delta I/I<5\%$). 
As discussed above, this sloping background is attributed to the scattering
due to single polarons \cite{xrayPolarons}, and to thermal-diffuse scattering.
A quantitative estimate of the thermal-diffuse contribution requires a full
calculation of the lattice dynamics that is beyond the scope of this paper.
Without this, we cannot make an accurate
estimate of the changes in the single polaron concentration in the applied 
field. However, the data of Fig. \ref{fig2}(b) suggest that the change in
the concentration of single polarons in the applied field is not significant,
which in turn implies
that single polarons play a less important role in the CMR effect than the
COO nanoclusters. Further experimental and theoretical work is needed to put
this suggestion on a quantitative basis.   

In conclusion, we report x-ray diffraction studies of the temperature-
and magnetic-field-dependent behavior of
charge/orbitally ordered nanoclusters 
in a single crystal of Nd$_{0.7}$Sr$_{0.3}$MnO$_3$. 
We find that the field-induced insulator-metal transition (the CMR effect)
in this compound, which is a typical CMR manganite, is accompanied by the
destruction of the COO nanoclusters.
The COO nanoclusters, however, do not completely disappear in the conducting
state, suggesting that this state is inhomogeneous and
always contains an admixture of the insulating phase.
We argue that these observations point to  
the important role of correlated lattice distortions in the CMR effect.

We are grateful to B. G. Kim for magnetic susceptibility measurements, and to
Y. J. Kim, A. J. Millis, and C. S. Nelson for important
discussions. This work was supported by the NSF under grants No. 
DMR-0093143, DMR-9802513, and by the DOE under contract No.
AC02-98CH10886.

%%%%%%%%%%%%%%%%%%%%%%%%%%%%%%%%%%%%%%%%%%%%%%%%%%%%%%%%%%%%%%%%%%%%%%%%

%%==============================================================================
%%==============================================================================

%%==============================================================================
%%==============================================================================

%%==============================================================================
%%==============================================================================


\begin{references}
%%%%%%%%%%%%%%%%%%%%%%%%%%%%%%%%%%%%%%%%%%%%%%%%%%%%%%%%%%%%%%%%%%%%%%%%

\bibitem{Review}
For a review, see {\it Colossal Magnetoresistance Oxides}, edited by 
Y. Tokura (Gordon \& Breach, London, 1999) 

\bibitem{Zener} C. Zener, Phys. Rev. {\bf 82}, 403 (1951); P. W. Anderson,
and H. Hasegawa, Phys. Rev. {\bf 100}, 675 (1955)

\bibitem{Millis} A. J. Millis, P. B. Littlewood, and B. I. Shraiman,
Phys Rev. Lett. {\bf 74}, 5144 (1995)

\bibitem{Brat} A. Alexandrov and A. M. Bratkovsky, Phys. Rev. Lett. {\bf 82}, 
141 (1999)

\bibitem{Polarons} S. J. L. Billinge, {\it et al}., Phys. Rev. Lett.
{\bf 77}, 715 (1996); C. H. Booth, {\it et al., ibid.} {\bf 80}, 853 (1998);
K. H. Kim, {\it at al., ibid.} {\bf 77}, 1877 (1996); M. Jaime, 
{\it et al., ibid.} {\bf 78}, 951 (1997); A. Machida,
{\it et al}., Phys. Rev. B {\bf 58}, R4281 (1998)

\bibitem{PDF} D. Louca, T. Egami, E. L. Brosha, H. R\"oder, and A. R. Bishop,
Phys. Rev. B {\bf 56}, R8475 (1997)

\bibitem{xrayPolarons} S. Shimomura, N. Wakabayashi, H. Kuwahara, and Y. Tokura,
Phys. Rev. Lett.
{\bf 83}, 4389 (1999); L. Vasiliu-Doloc, S. Rosenkranz, R. Osborn, S. K. Sinha,
J. W. Lynn, J. Mesot, O. H. Seeck,
G. Preosti, A. J. Fedro, and J. F. Mitchell, {\it ibid.}
{\bf 83}, 4393 (1999); 

\bibitem{D} for a review, see
E. Dagotto, T. Hotta, A. Moreo, Phys. Rep. {\bf 344}, 1 (2001)

\bibitem{Other}
M. Hennion, F. Moussa, G. Biotteau, and J. Rodriguez-Carvajal, 
Phys. Rev. B {\bf 61}, 9513 (2000) 

\bibitem{Corr} P. Dai, J. A. Fernandez-Baca, N. Wakabayashi, E. W. Plummer,
Y. Tomioka, and Y. Tokura, Phys. Rev. Lett. {\bf 85}, 2553 (2000);
C. P. Adams, J. W. Lynn, Y. M. Mukovskii, A. A. Arsenov, and D. A. Shulyatev,
{\it ibid.} {\bf 85}, 3954 (2000); 

\bibitem{Corr1} C. S. Nelson, M. v. Zimmermann, J. P. Hill, Doon Gibbs, 
V. Kiryukhin, T. Y. Koo, S-W. Cheong,
D. Casa, B. Keimer, Y. Tomioka, Y. Tokura, T. Gog, C. T. Venkataraman, 
Phys. Rev. B, in press 

\bibitem{Jaime} M. Jaime, P. Lin, S. H. Chun, M. B. Salamon, P. Dorsey,
and M. Rubinstein, Phys. Rev. B {\bf 60}, 1028 (1999)

\bibitem{DeT} J. M. DeTeresa, M. R. Ibarra, P. A. Algarabel, C. Ritter,
C. Marquina, J. Blasco, J. Garcia, A. del Moral, Z. Arnold,
Nature (London) {\bf 386}, 256 (1997)

\bibitem{Mydosh} M. Fath, S. Freisem, A. A. Menovsky,
Y. Tomioka, J. Aarts, and J. A. Mydosh, Science {\bf 285}, 1540 (1999)

\bibitem{KHKim} K. H. Kim, M. Uehara, and S-W. Cheong,
Phys. Rev. B {\bf 62}, R11945 (2000)

\bibitem{Chu} Y.-D. Chuang, A. D. Gromko, D. S. Dessau, T. Kimura, Y. Tokura,
Science {\bf 292}, 1509 (2001)

\end{references}
\end{document}